\input harvmac
\def\half{{1 \over 2}}
\def\dzm{{\partial}}

\def\p{{\partial}}
\def\s{{\sigma}}

\def\a {{\alpha}}
\def\b {{\beta}}

\def\g {{\gamma}}
\def\d {{\delta}}
\def\e {{\epsilon}}

\def\ad {{\dot\alpha}}
\def\bd {{\dot\beta}}

\def\t {{\theta}}
\def\ta {{\theta^\alpha}}

\def\tba {{\bar\theta^\ad}}

\def \ad {{\dot \a}}
\def \bd {{\dot \b}}
\def \t {{\theta}}
\def \tb {{\bar\theta}}

\Title{\vbox{\hbox{IFUSP-P-1211}}}
{\vbox{\centerline{\bf Super-Poincar\'e Invariant Koba-Nielsen Formulas
for the Superstring}}}
\bigskip\centerline{Nathan Berkovits}
\bigskip\centerline{Dept. de F\'{\i}sica Matem\'atica, Univ. de S\~ao Paulo}
\centerline{CP 20516, S\~ao Paulo, SP 01498, BRASIL}
\centerline{and}
\centerline{IMECC, Univ. de Campinas}
\centerline{CP 1170, Campinas, SP 13100, BRASIL}
\bigskip\centerline{e-mail: nberkovi@snfma2.if.usp.br}
\vskip .2in
The new spacetime-supersymmetric description of the superstring is
used to compute tree-level scattering amplitudes
for an arbitrary number of
massless four-dimensional states. The resulting Koba-Nielsen formula
is manifestly SO(3,1) super-Poincar\'e invariant and is easily generalized
to scattering in the presence of a D-brane.

\Date{April 1996}
\newsec {Introduction}

Before the work of this author, all spacetime-supersymmetric calculations of
superstring amplitudes used the light-cone Green-Schwarz formalism, which
requires non-trivial operators at the light-cone interaction points.\ref
\GS{M.B. Green and J.H. Schwarz, Nucl. Phys. B243 (1984) p.475.}
\ref\Mand{S. Mandelstam, Prog. Theor. Phys. Suppl. 86 (1986) p.163.}
\ref\RT{
A. Restuccia and J.G. Taylor, Phys. Rep. 174 (1989) p.283.}
Because
of this complication, only the four-point massless tree and one-loop
amplitudes
had been explicitly computed in a spacetime-supersymmetric manner.\foot{
Although reference \RT contains explicit expressions for 
four-point multiloop amplitudes,
these expressions contain unphysical divergences when interaction points
coincide. It has not yet been determined how these expressions are
affected by removing the divergences with contact terms.
Also, Mandelstam has constructed an N-point tree amplitude which
has manifest SU(4)$\times$U(1) super-Poincar\'e invariance.\ref
\Man2{S. Mandelstam, {\it Workshop on Unified String Theories,
29 July - 16 August 1985}, eds. M. Green and D. Gross, World
Scientific, Singapore (1986) p. 577.} However the
structure of his amplitude was obtained indirectly by unitarity
arguments, rather than by explicit computations.
It would be interesting to compare his amplitude with the Koba-Nielsen
formula in this letter.}

Over the last few years, this author has developed an alternative
spacetime-supersymmetric description of the superstring which has
the advantage of being conformally invariant, and therefore not requiring
non-trivial interaction-point operators.\ref\four{
N. Berkovits, Nucl. Phys. B431 (1994) p.258.}. 
The new formalism
is manifestly SO(3,1) super-Poincar\'e invariant and is suitable for any
compactification to four dimensions which preserves N=1 4D supersymmetry.
Furthermore, the new formalism contains critical N=2 worldsheet superconformal
invariance, and is related by a field redefinition to the N=2 embedding
of the N=1 RNS superstring.\ref\ust{N. Berkovits, Nucl. Phys. B420
(1994) p.332\semi N. Berkovits and C. Vafa, Mod. Phys. Lett. A9 (1994)
p.653.}

Together with Vafa, a simple ``topological'' method was developed
for calculating scattering amplitudes in this formalism.\ref\top
{N. Berkovits and C. Vafa, Nucl. Phys. B433 (1995) p.123.} (An easily
accessible review of this method can be found in reference \ref\me
{N. Berkovits, ``A New Description of the Superstring'', preprint
IFUSP-P-1212, April 1996, to appear on hep-th.})
In this letter, the topological method will be used to explicitly
compute tree-level scattering amplitudes for an arbitrary number of
four-dimensional massless states. The result can be expressed
as a Koba-Nielsen formula, and is manifestly SO(3,1) super-Poincar\'e
invariant.

It is straightforward to generalize the formula for scattering in the
presence of a D-brane. This result should be useful for D-brane
analysis since, unlike computations in the RNS formalism,\ref\Dbrane
{J. Polchinski, ``Dirichlet-Branes and Ramond-Ramond Charges'',
hep-th 9510017\semi S.S. Gubser, A. Hashimoto, I.R. Klebanov and
J.M. Maldacena, ``Gravitational Lensing by $p$-Branes'', hep-th
9601057\semi M.R. Garousi and R.C. Meyers, ``Superstring
Scattering from $D$-Branes'', hep-th 9603194.} it is
manifestly spacetime-supersymmetric, and therefore includes both
NS-NS and R-R couplings.

\newsec{Review}

As shown in reference \top, the topological
prescription for
computing the tree-level amplitude for N open superstring
states in any c=6 N=2 string theory is
\eqn\one{{\cal A}= < V_1(z_1)~ G^+(V_2(z_2))~
\tilde G^+(V_3(z_3)) \prod_{r=4}^N \int dz_r G^-(G^+(V_r(z_r)))>}
where $V(z)$ is a U(1)-neutral N=2 primary field of weight 0,
$[T,G^+,G^-,J]$ are the twisted c=6 N=2 generators (after twisting,
$G^+$ has weight 1 and $G^-$ has weight 2), $G^\pm(Y)$ means the contour
integral of $G^\pm$ around $Y$,  
$\tilde G^+ = G^- (e^{iH})$, and $J=\p H$.
The Chan-Paton factors will be supressed throughout this letter.

When \one is applied to the N=2 embedding of the N=1 RNS superstring,
it reproduces the standard RNS prescription in the ``large''
Hilbert space:
\eqn\two{{\cal A}= < \xi(z_1) V^{RNS}_1(z_1) ~Z V^{RNS}_2(z_2)
~V^{RNS}_3(z_3) \prod_{r=4}^N \int dz_r \{b,ZV^{RNS}_r(z_r)\}>}
since in the N=2 embedding, $V=\xi V^{RNS}$, $G^+ =j_{BRST}$, $G^- = b$,
and $\tilde G^+=\eta$. (Note
that $\beta=\p\xi e^{-\phi}$ and $\g=\eta e^{\phi}$
are the bosonic RNS ghosts, $Z=\{Q,\xi\}$ is the
RNS picture-changing operator, $V^{RNS}$ is 
in the
$-1$ picture, and
$ZV^{RNS}=\{Q,\xi V^{RNS}\}$
is in the 0 picture). 
So $G^+(V)=Q (\xi V^{RNS})= Z V^{RNS}$ and $\tilde G^+(V)=V^{RNS}$.

In the new spacetime-supersymmetric description of the superstring, a
field redefinition is used to write the generators of the N=2
embedding in terms of spacetime-supersymmetric variables.\four
In terms of the
four-dimensional supersymmetric
variables [$x^m,\t^\a,\tb^\ad,p_\a,\bar p_\ad,\rho$] and the 
compactification-dependent variables, 
the twisted c=6 N=2 generators are:
\eqn\gen{T=\half\p x^m \p x_m +
p_\a\p \t^\a + \bar p_\ad \p\tb^\ad +\half(\p\rho\p\rho+i\partial^2
\rho) +T_C,}
$$G^+=e^{i\rho} (d)^2 +G^+_C, \quad
G^-=e^{-i\rho} (\bar d)^2 +G^-_C, \quad
J=-\dzm\rho +J_C,
$$
where $m$ takes values 0 to 3, $\a$ and $\ad$ take values 1 or 2, 
$p_\a$ and $\bar p_\ad$ are conjugates to $\t^\a$ and $\tb^\ad$,
$\rho$ is a chiral boson,
$$d_\a=p_\a+{i\over 2}\tba\dzm x_{\a\ad}-{1\over 8}(\tb)^2\dzm\t_\a
+{1\over {16}}\t_\a \dzm (\tb)^2,$$
$$\bar d_\ad=\bar p_\ad
+{i\over 2}\ta\p x_{\a\ad}-{1\over 8}(\t)^2\p\tb_\ad
+{1\over {16}}\tb_\ad \dzm (\t)^2,$$
$(d)^2$ means
$\half\epsilon^{\a\b} d_\a d_\b$, 
and $[T_C,G_C^+,G_C^-,J_C]$ are the twisted c=9 N=2 generators of
the superconformal field theory
used to describe the compactification manifold.

The action in conformal gauge for these fields is
\eqn\act{\int dz d\bar z(\half\p x^m\bar\p x_m +p_{L\a}\bar\p\t_L^\a
+\bar p_{L\ad}\bar\p\tb_L^\ad +\half\p\rho_L\bar\p\rho_L}
$$ +p_{R\a}\p\t_R^\a
+\bar p_{R\ad}\p\tb_R^\ad +\half\p\rho_R\bar\p\rho_R)+S_C$$
where $S_C$ is the action for the compactification-dependent variables.
When the $R/L$ index has been suppressed (as in \gen), we shall always
mean left-moving.
Note that $z$ versus $\bar z$ is correlated with $L$ versus
$R$, and not with $\t$ versus $\tb$.

The free-field OPE's of the four-dimensional variables are
\eqn\ope{x^m(y) x^n(z)\to \eta^{mn}\log|y-z|^2,}
$$\rho_L(y) \rho_L(z) \to \log(y -z),
\quad \rho_R(y) \rho_R(z) \to \log(\bar y -\bar z),$$
$$p_{L\a}(y)\theta_L^\b (z)\to {\delta_\a^\b\over{y -z}},\quad
\bar p_{L\ad}(y)\bar\theta_L^\bd (z)\to {\delta_\ad^\bd\over{y -z}},$$
$$p_{R\a}(y)\theta_R^\b (z)\to {\delta_\a^\b\over{\bar y -\bar z}},\quad
\bar p_{R\ad}(y)\bar\theta_R^\bd (z)\to 
{\delta_\ad^\bd\over{\bar y -\bar z}}.$$
Note that the chiral boson $\rho$ can not
be fermionized since
$e^{i\rho(y)}~e^{-i\rho(z)}~\to 
(y -z)$. It has the same behavior as the
negative-energy field $\phi$ that appears when bosonizing the RNS ghosts.
After twisting, the background charge is defined for open
superstring tree amplitudes by
\eqn\norm{<(\t)^2 (\tb)^2 e^{-i\rho +iH_C}>=1}
where $J_C=\p H_C$.

As was shown by Siegel,\ref\sieg{W. Siegel, Nucl. Phys. B263 (1986) p.93.}
$d_\a$ and $\bar d_\ad$ satisfy
the OPE that $d_\a(y)$ $d_\b(z)$ is regular,
\eqn\dope{d_\a (y) \bar d_\ad(z) \to i{{\Pi_{\a\ad}}\over{y -z}}}
$$
d_\a(y) \Pi^m(z) \to -i{{\sigma^m_{\a\ad} \dzm\tb^\ad}\over{y -z}},\quad
\bar d_\ad(y) \Pi^m(z) \to -i{{\sigma^m_{\a\ad} \dzm\t^\a}\over{y -z}},$$
where
$$\Pi_{\a\ad}=\sigma^m_{\a\ad}
\dzm x_m -{i\over 2}\t_\a \dzm\tb_\ad-{i\over 2}\tb_\ad\dzm\t_\a.$$

The advantage of working with the variables $d_\a$ and $\Pi_m$ is that they
commute with the spacetime supersymmetry generators,
$$q_\a=\oint dz [p_\a -{i\over 2}
\tba\dzm x_{\a\ad}-{1\over {16}}(\tb)^2\dzm\t_\a],
\quad \bar q_\ad=\oint dz [\bar p_\ad
-{i\over 2}
\ta\dzm x_{\a\ad}-{1\over {16}}(\t)^2\dzm\tb_\ad].
$$

\newsec{Koba-Nielsen formula for open superstrings}

For massless states of the 
open superstring which are independent of the compactification, the
vertex operator
$V$ only depends on the zero modes of $x^m$, $\t^\a$ and $\tb^\ad$.
It can therefore be represented by the superfield $V(x,\t,\tb)$,
which is the standard N=1 4D super-Yang-Mills prepotential
(e.g. the $\t\tb$ component of $V$ is the gluon gauge field $A^m$, and the
$\t(\tb)^2$ and $(\t)^2\tb$ components of $V$ are the gluino fields
$\xi^\a$ and $\bar \xi^\ad$).

The condition of being N=2 primary implies that 
$$(D)^2 V = (\bar D)^2 V=\p_m \p^m V=0$$
where $D_\a=\p_{\t^\a}+{i\over 2}\tb^\ad\s^m_{\a\ad}\p_m$
and $\bar D_\ad=\p_{\tb^\ad}+{i\over 2}\t^\a\s^m_{\a\ad}\p_m$
are the covariant fermionic derivatives. Note that
$(D)^2 V = (\bar D)^2 V=0$
is the standard supersymmetric generalization of the Lorentz
gauge condition $\p_m A^m=0$.

Using the free-field OPE's of \ope, it is straightforward to compute that
$$\int dz G^-(G^+(V))=     
\int dz
(d^\a ~(\bar D)^2 D_\a V +
\bar d^\ad ~(D)^2\bar D_\ad V $$
$$+
\p\t^\a D_\a V 
-\dzm\tba~ \bar D_\ad V-i \Pi^{\a\ad}~[D_\a,\bar D_\ad] V ).$$

Plugging into \one, 
\eqn\first{{\cal A}= <V_1(z_1) e^{i\rho}d^\a(z_2) D_\a V_2(z_2)
e^{-2i\rho +H_C}\bar d^\ad(z_3) \bar D_\ad V_3(z_3) }
$$ \prod_{r=4}^N \int dz_r 
( d^\a(z_r) ~(\bar D)^2 D_\a V_r(z_r) 
+\bar d^\ad (z_r) ~(D)^2\bar D_\ad V_r(z_r) $$
$$+
\p\t^\a(z_r) D_\a V_r(z_r)
 -\dzm\tba(z_r)~ \bar D_\ad V_r(z_r)-i \Pi^{\a\ad}(z_r)
~[D_\a,\bar D_\ad] V_r(z_r) )>$$
$$=
(\prod_{r=1}^N {\p \over \p \e_r}|_{\e_r=0})
\prod_{s=4}^N\int dz_s ~\e_1
$$
$$
<\exp( \sum_{r=1}^N \e_r [
d_\a (z_r) w^\a_r 
+\bar d_\ad (z_r)\bar w^\ad_r
+\p\t_\a(z_r) a_r^\a
+\p\tb_\ad(z_r) \bar a_r^\ad +\Pi_{\a\ad}(z_r) a_r^{\a\ad}])$$
$$
e^{i\rho(z_2)} e^{-2i\rho(z_3)+iH_C(z_3)}\prod_{r=1}^N V_r(z_r)>$$
where
$$w_2^\a =D_2^\a,\quad \bar w_2^\ad=a_2^\a=\bar a_2^\ad=a_2^{\a\ad}=0,$$
$$\bar w_3^\ad =\bar D_3^\ad,\quad
w_3^\a=a_3^\a=\bar a_3^\ad=a_3^{\a\ad}=0,$$
$$\bar w_r^\ad =(D_r)^2\bar D_r^\ad,~ w_r^\a=
(\bar D_r)^2 D_r^\a,~ a_r^\a= D_r^\a,~\bar a_r^\ad=
- \bar D_r^\ad,~ a_r^{\a\ad}=-i [D_r^\a,\bar D_r^\ad]$$
for $r> 3$, and $D_r^\a$ and $\bar D_r^\ad$ are fermionic derivatives
which act only on $V_r(z_r)$. For example,
$$D_2^\a \prod_{r=1}^N V_r(z_r) =
V_1(z_1) D^\a V_2(z_2)\prod_{r=3}^N V_r(z_r).$$  

Using the fact that $<d^\a(z) ... >$=
$\sum_s <{{f^\a_s(z_s)}\over {z-z_s}}>$,
where $f_s$ is the residue of the pole at $z=z_s$, one can use the OPE's
of \dope to write
\eqn\sec{{\cal A}=
(\prod_{r=1}^N {\p \over \p \e_r}|_{\e_r=0})
\prod_{s=4}^N \int dz_s ~\e_1}
$$
<\exp( \sum_{r,s} \e_r
[{{D_{s\a}}
\over {z_r - z_s}} 
+\e_s({{i\Pi_{\a\ad}(z_s)\bar w^\ad_s -i\p\tb_\ad(z_s)
a_{s\a\ad}}\over 
{z_r - z_s}} +{{a_{s\a}}\over{(z_r - z_s)^2}})]w_r^\a$$
$$
+\half\sum_{r,s,t}\e_r\e_s\e_t {{\p\tb_\ad(z_s)\bar w^\ad_s w_{t\a}w^\a_r
}\over{(z_r - z_s)(z_t- z_s)}} 
$$
$$+\sum_r \e_r[\bar d^\ad (z_r)\bar w^\ad_r
+\p\tb_\ad(z_r) \bar a_r^\ad +\Pi_{\a\ad}(z_r) a_r^{\a\ad}])$$
$$
e^{i\rho(z_2)} e^{-2i\rho(z_3)+H_C(z_3)}\prod_{r=1}^N V_r(z_r)>$$
where 
$D_r^\a$ is always ordered to the left of $w_r$ and $a_r$.
The term proportional to $\e_r\e_s\e_t$
comes from the pole of $d^\a(z_t)$ with the residue of the pole
of $d^\b(z_r)$ at $z_s$.

One can similarly use the OPE's for $\bar d^\ad$ to write 
\eqn\third{{\cal A}=
(\prod_{r=1}^N {\p \over \p \e_r}|_{\e_r=0})
\prod_{s=4}^N \int dz_s  ~\e_1}
$$
<\exp( \sum_{r,s} \e_r 
[{{D_{s\a}w_r^\a +\bar D_{s\ad}\bar w_r^\ad}\over {z_r - z_s}}
+\e_s({{i\Pi_{\a\ad}(z_s)\bar w^\ad_s w^\a_r
}\over 
{z_r - z_s}} +{{a_{s\a} w_r^\a
+\bar a_{s\ad}\bar w_r^\ad}\over{(z_r - z_s)^2}})]
$$
$$-i
\sum_{r,s,t}\e_r\e_s\e_t 
{{ a_{t\a\ad} w^\a_s \bar w_r^\ad}
\over 
{(z_r - z_t)^2(z_s - z_t)}}$$
$$+\half
\sum_{r,s,t,u}
\e_r\e_s\e_t\e_u {{\bar w_{t\ad} w_{u\a} w^\a_s \bar w_r^\ad}\over
{(z_r - z_t)^2(z_s -z_t)
(z_u -z_t)}}
 +\sum_r \e_r\Pi_{\a\ad}(z_r) a_r^{\a\ad})$$
$$
e^{i\rho(z_2)} e^{-2i\rho(z_3)+iH_C(z_3)}\prod_{r=1}^N V_r(z_r)>$$
where $\bar D_r^\ad$ is always ordered to the left of
$D_r^\a$, which is always ordered to the left of $w_r$ and $a_r$.

Finally, performing the correlation function over the $x$'s and $\rho$,
and using the normalization of \norm, one obtains
\eqn\open{{\cal A}=
(\prod_{r=1}^N {\p \over \p \e_r}|_{\e_r=0})
\prod_{s=4}^N \int dz_s  ~\e_1}
$$
\int d^2\t_0 d^2\tb_0 :\exp( \sum_{r,s} \e_r 
[{{D_{s\a}w^\a_r +\bar D_{s\ad}\bar w^\ad_r 
+2k_{sm}a_r^m }\over {z_r - z_s}}
+\e_s{{a_{s\a}w_r^\a +\bar a_{s\ad}\bar w_r^\ad}\over{(z_r - z_s)^2}}]
$$
$$+i\sum_{r,s,t}\e_r\e_s({ {
2 w^\a_r \bar w^\ad_s
k_{t\a\ad}}
\over 
{(z_s - z_r)(z_s -z_t)}} 
+ \e_t {{w^\a_r \bar w^\ad_s a_{t\a\ad}}\over {(z_s -z_t)(z_r-z_t)
(z_r - z_s)}})
$$
$$+{1\over 4}\sum_{r,s,t,u}
\e_r\e_s\e_t \e_u {{\bar w_{s\ad} \bar w_u^\ad w_{t\a}w_r^\a}
\over {(z_r - z_s)(z_t - z_u)(z_r - z_u)(z_t -z_s)}}):$$ 
$$
(z_2 - z_3)^2 \prod_{r,s} (z_r -z_s)^{4 k_r^m k_{sm}}\prod_r
V_r(z_r)$$
where
$k_r^m$ is the momentum of $V_r$,
$:\exp():$ is to remind that
$\bar D_r^\ad$ is always ordered to the left of
$D_r^\a$ which is always ordered to the left of $w_r$ and $a_r$,
and $\int d^2\t_0
d^2\tb_0$ means
to set $\t^\a(z_r)=\t_0^\a$ and $\tb^\ad(z_r)=\tb^\ad_0$ for all $r$
and to integrate over $\t_0^\a$ and $\tb_0^\ad$.

It is straightforward to check that ${\cal A}$ is gauge-invariant
since under the 
transformation 
$\d V_p=(\bar D)^2 \Phi_p$ 
(which implies $a_p^\ad= D_p^\ad$, $a_p^m= k_p^m$,
and $\bar a_p^\ad=w_p^\a$
=$\bar w_p^\ad=0$)
\eqn\gauge{\d{\cal A}=
(\prod_{r=1}^N {\p \over \p \e_r}|_{\e_r=0})
\prod_{s=4}^N \int dz_s  ~\e_1\e_p {\p \over{\p z_p}}}
$$
\int d^2\t_{0} d^2\tb_{0} 
:\exp F:
(z_2 - z_3)^2 \prod_{r,s} (z_r -z_s)^{k_r^m k_{sm}}
(\bar D_p)^2 \Phi_p\prod_{r\neq p}
V_r(z_r)$$
where
$F$ is the object in the exponential of \open.

\newsec{Koba-Nielsen formula for closed superstrings}

For the closed superstring, the tree-level scattering amplitude of
compactification independent massless states is obtained by
multiplying ${\cal A}$ by its right-moving counterpart and by replacing
the 4D N=1 super-Yang-Mills prepotential $V(x,\t,\tb)$ with
the 4D N=2 supergravity plus tensor prepotential 
$U(x,\t_L,\tb_L;$
$\t_R,\tb_R)$.
In standard SU(2) notation for N=2 superspace,
$\t_L^\a=\t_+^\a$, 
$\tb_L^\ad=\tb^{+\ad}$, 
$\t_R^\a=\t_-^\a$, and
$\tb_R^\ad=\tb^{-\ad}$. 

As discussed in reference \ref\mesieg
{N. Berkovits and W. Siegel,
``Superspace effective actions for 4D compactifications
of heterotic and type II superstrings'',
preprint IFUSP-P-1180, October 1995, to appear in Nucl. Phys. B,
hep-th 9501016.}, $U$ is a scalar superfield which describes
an N=2 conformal supergravity multiplet and an N=2 tensor hypermultiplet.
The NS-NS fields for the graviton, anti-symmetric
tensor, and dilaton are in the $\t_R\tb_R\t_L\tb_L$ component of $U$,
$$h_{mn}+b_{mn}+\eta_{mn}\phi=\s_m^{\a\ad}\s_n^{\b\bd}
[D_{L\a},\bar D_{L\ad}]
[D_{R\b},\bar D_{R\bd}] U,$$
and the R-R field strengths
for the U(1) vector 
and complex scalar are in the 
$(\t_L)^2\tb_L$
$ (\t_R)^2\tb_R$
and $\t_L(\tb_L)^2 $
$(\t_R)^2\tb_R$ components of $U$, 
$$F_{mn}=\s_{mn}^{\ad\bd}
(D_L)^2 \bar D_{L\ad}(D_R)^2\bar D_{R\bd}U+ c.c.,\quad
\p_m y=\s_m^{\a\bd}D_{L\a}(\bar D_L)^2(D_R)^2\bar D_{R\bd}U.$$

For $U$ to be an N=2 primary field, it must satisfy the constraints
$$(D_L)^2 U=(\bar D_L)^2 U=(D_R)^2 U=(\bar D_R)^2 U=\p_m \p^m U=0.$$
The first four constraints are the N=2 supersymmetric generalization of the
usual polarization conditions, and the last constraint is the equation of
motion in this gauge.

The topological prescription for the N=2 closed superstring tree-level
amplitude is
$${\cal A}=
< U_1(z_1) ~ G_R^+( G_L^+(U_2(z_2)))
~\tilde G_R^+(\tilde G_L^+(U_3(z_3))) $$
$$\prod_{r=4}^N \int dz_r d\bar z_r
G_R^-(G_L^-( G_R^+(G_L^+(U_r(z_r)))))>$$
where $G_R^\pm$ and $G_L^\pm$
are the right and left-moving fermionic N=2 generators.

Using the same methods as in the open superstring computation, one finds
\eqn\closed{{\cal A}=
(\prod_{r=1}^N 
{\p \over \p \e_{Lr}}{\p\over\p\e_{Rr}}|_{\e_{Lr}=e_{Rr}=0})
\prod_{s=4}^N \int dz_s d\bar z_s
~\e_{L1}\e_{R1}}
$$
\int d^2\t_{L0} d^2\tb_{L0} 
\int d^2\t_{R0} d^2\tb_{R0} :\exp(F_L +F_R):
$$
$$|z_2 - z_3|^4 \prod_{r,s} |z_r -z_s|^{2k_r^m k_{sm}}\prod_r
U_r(z_r)$$
where
$F_L$ is the object in the exponential of \open with $k_r^m$ replaced
by $\half k_r^m$, and 
$F_R$ is its right-moving counterpart (i.e. $F_R$
is obtained from $F_L$ by switching $(\e_{Lr},D_{Lr}, w_{Lr}, a_{Lr})$
to $(\e_{Rr},D_{Rr}, w_{Rr}, a_{Rr})$
and by switching $z_r$ to $\bar z_r$).

\newsec{Koba-Nielsen formula in the presence of a $D$-brane}

For closed string scattering in the presence of a D-brane, there are
boundary conditions on the fields which can be either Neumann or
Dirichlet.
For convenience, the boundary will be chosen
to be the real line,
and the closed strings to reside in the upper half-plane.

The boundary conditions
when $z=\bar z$ for the four-dimensional fields are\ref\Vipul{
V. Periwal and \O. Tafjord, private communication\semi
M.B. Green, Phys. Lett. B329 (1994), p. 435.}
\eqn\boundary{\p x^m =(-1)^{n_m} \bar\p x^m,\quad
\t_L^\a = M^\a_{\,\b}\t_R^\b,\quad
d_L^\a = M^\a_{\,\b}d_R^\b,} 
$$\tb_L^\ad = M^\ad_{\,\bd}\tb_R^\bd,\quad
\bar d_L^\ad = M^\ad_{\,\bd} \bar d_R^\bd, 
\quad \rho_L=\rho_R$$
where $M^\a_{\,\b}$ and 
$M^\ad_{\,\bd}$ are the matrix elements of 
$M=\prod_{m=0}^3 \g_m^{n_m}$, and 
$n_m=0$ or 1 for Neumann or Dirichlet boundary conditions in the
$m^{th}$ spacetime direction. (We are assuming in this letter that
$\sum_{m=0}^3 n_m$ is even. When $\sum_{m=0}^3 n_m$ is odd, 
one needs to switch
$\t_R^\a$ with $\tb_R^\ad$ and $d_R^\a$ with $\bar d_R^\ad$.)
Note that because the four-dimensional variables are independent of
the compactification, \boundary is independent of the boundary conditions
in the compactification directions.

The first step in calculating the scattering with a D-brane is to
write down the topological prescription. Comparing with the RNS 
prescription,\Dbrane
\eqn\Dpres{{\cal A}= < \xi_L(z_1) V^{RNS}_1(z_1)\int dz_2 \{b_L,
Z_R Z_L V^{RNS}_2(z_2)\} }
$$
\prod_{r=3}^N \int dz_r d\bar z_r [b_R,\{b_L,Z_R Z_L V^{RNS}_r(z_r)\}]>,$$
and using the N=1/N=2 relationship which is discussed after \two, one sees
that the correct topological prescription is
\eqn\toppres{{\cal A}=
< \tilde G^+_R(U_1(z_1))\int dz_2 G^-_L (G^+_L(
 G^+_R(U_2(z_2)))) }
$$
\prod_{r=3}^N \int dz_r 
d\bar z_r G_R^-(G_L^-( G_R^+(G_L^+(U_r(z_r)))))>.$$

Plugging in the vertex operator for the massless fields,
\eqn\Dcalc{{\cal A}=
(\prod_{r=1}^N {\p \over \p \e_{Lr}}
{\p \over \p \e_{Rr}}
|_{\e_{Lr}=\e_{Rr}=0})
\int dz_2\prod_{s=3}^N \int dz_s d\bar z_s\e_{L1}}
$$
<\exp( \sum_r \e_{Lr} (
d_{L\a} (z_r) w^\a_{Lr} 
+\bar d_{L\ad} (z_r) \bar w^\ad_{Lr} $$
$$
+\p\t_{L\a}(z_r) a_{Lr}^\a
 +\p\tb_{L\ad}(z_r) \bar a_{Lr}^\ad +\Pi_{Lm}(z_r) a_{Lr}^{m})$$
$$\exp( \sum_r \e_{Rr} (
d_{R\a}(\bar z_r) w^\a_{Rr} 
+\bar d_{R\ad} (\bar z_r) \bar w^\ad_{Rr} $$
$$
+\bar\p\t_{R\a}(\bar z_r) a_{Rr}^\a
 +\p\tb_{R\ad}(z_r) \bar a_{Rr}^\ad +
\Pi_{Rm}(\bar z_r) a_{Rr}^{m})$$
$$
e^{-2i\rho_R(\bar z_1)+iH_{RC}(\bar z_1)}
e^{i\rho_R (\bar z_2)}\prod_{r=1}^N U_r(z_r)>$$
where the only non-zero $w$'s and $a$'s are
$$\bar w_{r}^\ad =(D_{r})^2\bar D_{r}^\ad,~ w_{r}^\a=
(\bar D_{r})^2 D_{r}^\a,~ a_{r}^\a= D_{r}^\a,~\bar a_{r}^\ad=
-\bar D_{r}^\ad,~ a_{r}^{\a\ad}=-i[D_{r}^\a,\bar D_{r}^\ad]$$
for $r>1$ when left-moving and for $r>2$ when right-moving, and
$$\bar w_{R1}^\ad =\bar D_{R1}^\ad, w_{R2}^\a= D_{R2}^\a.$$

One can now use the techniques of section 3 to compute the
correlation function. The only difference is that
the correlation 
function of left-moving fields must equal the correlation
function of its right-moving partner 
when the field sits on
the boundary.
For example,
$$<d_L^\a (z) ... >=< \sum_s {{f_{Ls}^\a (z_s)}\over{z-z_s}}
+\sum_t {{f_{Rt}^\a (z_t)}\over{z-\bar z_t}}>$$
where $f_{Ls}^\a$ is the pole residue of $d_L^\a(z)$ at $z=z_s$
and $f_{Rt}^\a$ is the pole residue of $M^\a_{\,\b} d_R^\b(z)$ at 
$\bar z=\bar z_t$.

The resulting Koba-Nielsen formula in the presence of a D-brane
is:
\eqn\Dbrane{{\cal A}=
(\prod_{r=1}^N 
{\p \over \p \e_{Lr}}{\p\over\p\e_{Rr}}|_{\e_{Lr}=e_{Rr}=0})
\int dz_2 \prod_{s=3}^N \int dz_s d\bar z_s~\e_{L1}
\int d^2\t_{0} d^2\tb_{0}} 
$$
:\exp[~[~\sum_{r,s} \e_{Lr}  
({{D_{Ls\a}w_{Lr}^\a +\bar D_{Ls\ad} 
\bar w_{Lr}^\ad 
+k_{sm}a_{Lr}^m }\over {z_r -  z_s}}
+\e_{Ls}{{a_{Ls\a}w_{Lr}^\a +
\bar a_{Ls\ad}\bar w_{Lr}^\ad}\over{(z_r - z_s)^2}})
$$
$$+i [~\sum_{r,s,t}\e_{Lr}\e_{Ls}(
{{w_{Lr}^\a \bar w^\ad_{Ls}
k_{t\a\ad}}
\over 
{(z_s - z_r)(z_s - z_t)}} 
+ \e_{Lt} {{w^\a_{Lr} 
\bar w^\ad_{Ls} 
a_{Lt\a\ad}}\over {( z_s - z_t)(z_r- z_t)
(z_r -  z_s)}}) $$
$$+{1\over 4}[~\sum_{r,s,t,u}
\e_{Lr}\e_{Ls}\e_{Lt} \e_{Lu}
 {{\bar w_{Ls\ad} \bar w_{Lu}^\ad w_{Lt\a}w_{Lr}^\a}
\over {(z_r - z_s)(z_t - z_u)(z_r - z_u)(z_t -z_s)}} $$
$$+ \{Lu\to Ru\} ~]+
\{Lt \to Rt\}~] + \{Ls \to Rs\} ~]+\{Lr \to Rr\}~]$$
$$(\bar z_1 - \bar z_2)^2 \prod_{r,s} |z_r -z_s|^{2k_r^m k_{sm}}
|z_r -\bar z_s|^{2 k_r^m \hat k_{sm}}\prod_r
U_r(z_r)$$
where
$\hat k_r^m = (-1)^{n_m} k_r^m$,
$\int d^2\t_0 d^2\tb_0$ means to set
$\t^\a_L(z_r)=M^\a_{\,\b} \t_R^\b (\bar z_r)=\t_0^\a$
and
$\bar\t^\ad_L(z_r)=M^\ad_{\,\bd} \tb_R^\bd (\bar z_r)=\tb_0^\ad$
for all $r$ and to integrate over $\t_0^\a$ and $\tb_0^\ad$, and
$\{Lu\to Ru\}$ means to switch $z_u \to \bar z_u$ and to switch
$$
(\e_{Lu}, ~w_{Lu}^\a,~ \bar w_{Lu}^\ad,~ a_{Lu}^\a,~ a_{Lu}^\ad,~a_{Lu}^m,
~k_u^m)\to $$
$$
(\e_{Ru},~M^\a_{\,\b} w_{Ru}^\b, 
~M^{\ad}_{\,\bd}\bar w_{Lu}^\bd,~ M^{\a}_{\,\b}a_{Lu}^\a, 
~M^{\ad}_{\,\bd}a_{Lu}^\bd,~(-1)^{n_m}a_{Lu}^m,~\hat k_u^m).$$
Note that the switch affects everything inside the $[~]$ brackets, e.g.
$\{Lu\to Ru\}$ affects only the $\e_r\e_s\e_t\e_u$ term but
$\{Lr \to Rr\}$ affects all terms (including those coming from
earlier switches).

In summary, the manifestly super-Poincar\'e invariant
Koba-Nielsen formulas for massless
tree amplitudes of the open superstring,
closed superstring, and closed superstring in the presence of a 
D-brane can be found in equations \open, \closed, and \Dbrane.

Acknowledgements: I would like to thank Vipul Periwal and
\O jvind Tafjord
for suggesting
the D-brane generalization and for sharing their results using
light-cone Green-Schwarz calculations.
\listrefs
\end